\documentclass[12pt,aps,prb,titlepage]{article}
\usepackage{setspace}
\usepackage{epsfig}
\usepackage{ae}
\usepackage{aecompl}
\usepackage{amsmath}
\usepackage{graphicx}
\usepackage[round,numbers,sort&compress]{natbib}
\begin{document}

\title{Dependence of the energies of fusion on the 
inter-membrane separation: optimal and constrained}

\author{J.Y. Lee and M. Schick\\
Department of Physics\\ University of Washington,  Box
  351560, Seattle, WA 98195-1560}

\date{} 
\maketitle

\begin{abstract} 
We calculate the characteristic energies of fusion between planar bilayers as a
function 
of the distance between them, measured from the hydrophobic/hydrophilic 
interface of one of the  two nearest, cis, leaves to the other.
The two  leaves of each bilayer are of equal composition;  0.6 volume
fraction of a lamellar-forming amphiphile, such as
dioleoylphosphatidylcholine, and 0.4 volume fraction of a
hexagonal-forming amphiphile, such as
dioleoylphosphatidylethanolamine.  
Self-consistent field theory is employed to solve the model. We
find that the largest barrier to fusion is that to create the metastable
stalk. This barrier is the smallest, about 14.6 $k_BT$,  when the bilayers
are at a distance about 20 percent greater than the thickness of a
single leaf, a distance which would correspond to between two and three
nanometers for typical bilayers. The very size of the protein machinery
which brings the membranes together can prevent them from reaching this
optimum separation. For even modestly larger separations, 
we find a linear rate of increase of the free energy
with distance between bilayers for the metastable stalk
itself and for the barrier to the creation of this stalk. We estimate
these rates for biological membranes to be  about 7.1 $k_BT$/nm and 16.7
$k_BT$/nm respectively. The major contribution to this rate 
comes from the increased packing energy
associated with the hydrophobic tails. From this we estimate, for the
case of hemagglutinin, a free energy of 38 $k_BT$ for the metastable
stalk itself, and a barrier to create it of 73 $k_BT$. Such a large
barrier would require that more than a single hemagglutinin molecule be
involved in the fusion process, as is observed.  
\end{abstract}
\section{Introduction}
Although it is essential to a host of biological processes in which material 
enters, exits, or changes location within the cell, ({\em e.g.} viral 
entry, exocytosis, and intracellular trafficking) the process of membrane 
fusion is not well understood. Some basic concepts, however, are clear. 
The membranes to be fused must be put under tension, {\em i.e.} their free 
energy per unit area must be increased, so that the fused state with smaller area
has a lower free energy than the unfused system.   This tension is brought about
by bringing  the membranes to be fused in close proximity 
to one another,  on the order of a few nanometers, thereby removing  
some water  from 
the hydrophilic headgroups of the amphiphiles  comprising the 
membrane and consequently raising the system free energy.
This additional energy is supplied by fusion proteins.
Even though the  free energy of the system is reduced by fusion, 
the rearrangement of lipids 
required by the process can only occur if the system surmounts free
energy barriers.  The
calculation of these barriers has been the subject of much attention 
\cite{Kozlov83,Siegel93,Kuzmin01,Lentz00,Chernomordik03}. 

From the above argument, it follows that the
 barrier to the fusion process must be a function of the 
tension. 
It also depends on the pathway to fusion that the system takes
\cite{Katsov04,Katsov06}, as well as  several other factors. Among these are the average
compositions of the different amphiphiles comprising the membrane and,
in particular, their composition in the cis, or proximal, 
leaves \cite{Lee07,Kasson07}. We have examined each of these factors, and
the upshot is, that for bilayers in which the relative fraction of
hexagonal-forming and lamellar-forming amphiphiles in the cis leaves are
similar to that in biological membranes, the largest barrier to fusion,
in either the standard \cite{Kozlov83} or non-standard 
\cite{Noguchi01,Mueller02,Mueller03} pathways, 
is that to form the initial stalk. This is the initial local junction 
formed by the rearrangement of lipids in the two apposing cis leaves \cite{Kozlov83}. 
Further, this barrier is not large;
it was estimated \cite{Lee07} to be on the order of 13$k_BT$, with $T$ the absolute
temperature and $k_B$ Boltzmann's constant. That the 
rate-limiting barrier to fusion should be so small led us to conclude that 
fusion should occur rapidly once the two membranes were brought 
sufficiently close to initiate the process.

This conclusion highlights the question of what is ``sufficiently
close'', {\em i.e.} the
issue of the dependence of the fusion 
barrier on the distance between the membranes to be fused. It is an 
interesting issue which speaks to the interplay between the lipids and a fusion 
protein. An example is provided by 
hemagglutinin, the fusion protein associated with the influenza 
virus \cite{Eckert01}.
It is anchored in the viral membrane. A
cluster of between three and six of them around the eventual site of fusion 
are required \cite{Blumenthal96}.  
A first conformational change of hemagglutinin 
is accompanied by removal of 
the receptor binding domains. A second conformational change 
exposes the hydrophobic fusion 
peptide which anchors in the target membrane.  
At this point the conformation of the several hemagglutinins, which are
essentially normal to the membranes, keep the 
viral and target membranes at a distance 
of 13.5 nm from one another \cite{Wiley87}. A final conformational
change brings the 
membranes much closer, on the order of 4 nm,  with the hemagglutinin 
now parallel to the membranes and pointing away from the fusion site 
\cite{Bentz93a}. This conformational
change releases a great deal of energy, on the order of 60$k_BT$
per hemagglutinin \cite{Kozlov98}, which presumably is expended in
pulling the membranes to this distance and in bringing about 
the formation of the
stalk.  
The question is why this distance is what it is. Is it because a
smaller distance between membranes would cause fusion to be
energetically less expensive, but the very size of the
hemagglutinin prevents a closer approach, or is it that the machinery
is such that it does bring 
the membranes to the optimal separation?
Just what is the competition 
that sets the distance at which fusion occurs?
Similar questions apply to the SNARE
machinery which promotes fusion 
\cite{Sollner93}.

There has been little theoretical work on the distance dependence of the 
barrier to fusion \cite{Kozlovsky02,Knecht03,Kasson06}. 
It was considered explicitly by Kozlovsky and Kozlov 
\cite{Kozlovsky02} using a phenomenological model. They found that 
the energy of an isolated stalk was 
practically independent of the distance between membranes, and approached 
a value of about 43$k_BT$ as the distance between membranes increased  
without limit. This result can be traced to a few assumptions. First the 
membranes are assumed to be tensionless. Hence, the additional membrane 
area needed to create a stalk between two membranes at a large distance
costs no free 
energy by assumption. This assumption is presumably quite good when the
distance between membranes is greater than that of the hydrophobic
repulsion, on the order of a few nm \cite{Israelachvili82}. The second 
assumption is that the
membranes can bend to take a shape which minimizes the curvature energy
of the system. Given the constraints on the membrane separation placed
by the presence of the fusion proteins, this is probably not the case.
Finally, the phenomenological free energy employed does not capture the
energy associated with packing the tails efficiently into the axially
symmetric stalk structure, a structure very different from the planar
bilayer, the membrane configuration of lowest free energy. 

In order to clarify these issues, particularly that of the packing, 
we employ a microscopic model to study  
the dependence of the barriers to fusion on the 
distance, $H$, between the hydrophobic/hydrophilic interfaces of the
apposed leaves of planar membranes. The membranes are under either zero
or a small tension. The 
membranes are composed of a mixture of two amphiphiles, one lamellar- and 
the other hexagonal-forming. The leaves are of equal composition, one that 
mimics the mix of these two classes of amphiphiles in the cis leaves of 
red blood cell membranes. This choice is made because previous work 
\cite{Kozlovsky02,Kasson07} shows that the free energy of fusion 
intermediates is 
most sensitive to the composition of the cis leaves, and rather 
insensitive to that of the trans leaves.  
Only the standard fusion mechanism is considered. We do 
this because we have found very little difference in the 
barriers of the two different mechanisms when membranes with a mix of 
hexagonal and lamellar formers were considered \cite{Lee07}. In addition, this 
restriction significantly simplifies the calculation.

We find, once again, that the largest barrier to fusion is  that 
associated with the formation of the initial stalk. We also can
understand the dependence of this barrier on the separation between
membranes as
follows. When the membranes are very close, the barrier to fusion
increases with decreasing distance for two reasons. Not only does the
repulsive, hydrophobic, interaction, essentially a
depletion force, increase with decreasing separation, 
but also the energy required for the amphiphiles to
rearrange into a stalk of such short extent becomes larger with smaller
membrane separation. Due to this effect, the stalk is not a metastable
structure. As a consequence  fusion would have to proceed directly to a fusion
pore without a stalk intermediate, an absence which would  make 
the process much less
likely. When the membranes are farther apart, the stalk becomes a stable
intermediate, and the barrier to fusion decreases. As the distance
between membranes increases still further, the barrier to
fusion now increases rapidly with increasing distance due to the  
packing energy of
the initial stalk connecting the membranes, an energy  which scales with
the length of the stalk. We find this rate of increase to be about 
7 $k_BT$ per nm.
Consequently the lowest barrier to fusion occurs when the two membranes
are at a distance large enough that membrane repulsion is not too great,
and the stalk is metastable, but small enough that the stalk is
relatively short and energetically inexpensive. In our system we find
the optimum distance to be
about twenty percent greater than 
the thickness of a single leaf of our bilayer, a distance which would 
correspond to between two and three nanometers for typical
membranes. This is in reasonable agreement with the observed distance to which
laboratory membranes must be brought in order to fuse \cite{Weinreb07}. 
The lowest barrier to fusion corresponds to about
$14.6k_BT$ for a biological membrane. To fuse membranes which are at a
somewhat larger distance, as in the case when the very size of
hemagglutinin prevents a closer approach, requires traversing a
larger barrier. At a distance between headgroups of 4 nm 
applicable to the case of
hemagglutinin, we estimate that the barrier is on the order of 73$k_BT$. 
It is not surprising, then, that 
more than a single fusion protein would be required.
 
 \section{The model}
 To investigate the effect of the distance between planar membranes on the 
free energy barrier to fuse them, we extend the application of 
self-consistent field theory to microscopic models of membranes initiated 
earlier \cite{Katsov04,Katsov06,Lee07}. The basic assumption of this 
approach is that the self-assembly into bilayer vesicles and the processes 
which these vesicles can undergo, such as fusion, are common to systems of 
amphiphiles, of which lipids are but one example. Recent work on vesicles 
which consist of diblock copolymers serves to illustrate this point 
\cite{Discher99}. It follows that these processes can be explored in 
whatever system of amphiphiles proves to be most convenient. For the 
application of self-consistent field theory, that system is one of block 
copolymers in a homopolymer solvent. While the processes that amphiphiles 
undergo are presumably universal, the energy scales of these processes are 
system-dependent, and thus it is necessary to be able to compare the 
energy scale in a biological bilayer with the energy scale in our system of block 
copolymers. This will be done below.

Here we consider a system of two bilayers each composed of two different 
amphiphiles that resemble 
dioleoylphosphatidylcholine, (DOPC), and 
dioleoylphosphatidylethanolamine, (DOPE),
in their hydrophobic/hydrophilic 
ratios. The two leaves of each bilayer are of the same composition. The 
system is incompressible and occupies a volume $V$. The two amphiphiles 
are each AB diblock copolymers. Type 1, a lamellar-former, consists of $N$ 
monomers and has a molecular volume $Nv$. The fraction of hydrophilic 
monomers, arbitrarily chosen to be of type $A$, is denoted $f_1$ and is 
assigned the value $f_1=0.4$ as such a diblock has a ``spontaneous 
curvature" similar to that of DOPC \cite{Katsov04}.  The amphiphile of 
type 2 consists of $N\tilde\alpha$ monomers and has a molecular volume of 
$\tilde\alpha Nv$. The fraction of hydrophilic monomers, $f_2$ is chosen 
to be $f_2=0.294$ as this produces a spontaneous curvature similar to that 
of DOPE. We set $(1-f_1)Nv=(1-f_2)\tilde{\alpha} Nv$ such that hydrophobic 
tails of different types of amphiphiles have equal length. For our chosen 
$f_1=0.4$ and $f_2=0.294$, $\tilde\alpha = 0.85$. The solvent is an A 
homopolymer with volume $Nv$.

We denote the local volume fraction of hydrophilic elements of
amphiphile 1 to be $\phi_{A,1}({\bf r})$, of amphiphile 2 to be 
$\phi_{A,2}({\bf r})$, and
of the solvent to be $\phi_{A,s}({\bf r})$. The total local volume
fraction of hydrophilic elements is denoted 
\begin{equation}
\phi_A({\bf r})=\phi_{A,1}({\bf r})+\phi_{A,2}({\bf r})+\phi_{A,s}({\bf
r}).
\end{equation}
Similarly the total local volume fraction of hydrophobic elements is
\begin{equation}
\phi_B({\bf r})=\phi_{B,1}({\bf r})+\phi_{B,2}({\bf r}).
\end{equation}
The 
amounts of each of the components are controlled by activities,
$\zeta_1$, $\zeta_2$, and $\zeta_s$.   
Because of the incompressibility constraint, only two of
the activities are independent.  
Cylindrical coordinates, $(\rho,\theta,z)$, are employed.

Within the self-consistent field approximation, the free energy, 
$\Omega(T,V,{\cal A},\zeta_1,\zeta_2,\zeta_s)$, of the system containing a
bilayer, or bilayers,  each of area 
${\cal A}$, is given by the minimum of the functional
\begin{eqnarray}
\label{free1}
\frac{Nv}{k_BT}{\tilde\Omega}&=&-\zeta_1Q_1-\zeta_2Q_2-\zeta_sQ_s\nonumber \\
                  &&+\int d{\bf r}[\chi N\phi_A({\bf r})\phi_B({\bf r})-
w_A({\bf r})\phi_A({\bf r})-w_B({\bf r})\phi_B({\bf r})\nonumber \\
                  &&-\xi({\bf r})(1-\phi_A({\bf r})-\phi_B({\bf r}))],
\end{eqnarray} 
where $Q_1(T,[w_A,w_B])$, $Q_2(T,[w_A,w_B]),$ and $Q_s(T,[w_A])$ are the
configurational parts of the single chain partition functions of amphiphiles 
1 and 2 and of
solvent. They have the dimensions of volume, and 
are functions of the temperature, $T$, which is inversely
related to the Flory interaction $\chi$, and functionals of
the fields $w_A$ and $w_B$. 
These fields, and the Lagrange multiplier
$\xi({\bf r})$, which enforces the local incompressibility condition, are
determined by the self-consistent equations which result from minimizing 
the free energy functional.
Insertion of these fields into the free energy functional, Eq. (\ref{free1}), 
yields the free energy within the self-consistent field approximation:
 \begin{eqnarray}
 \label{scsymfree}
  \frac{Nv}{k_BT}\Omega(T,V,{\cal A},\zeta_1,\zeta_2,\zeta_s)
&=&-\zeta_1Q_1(T,[w_A,w_B])-\zeta_2Q_2(T,[w_A,w_B])-\zeta_sQ_s(T,[w_A])\nonumber\\
  &-&\int\ d{\bf r}\chi N\phi_A({\bf r})\phi_B({\bf r}),
\end{eqnarray}
The free energy of the system without the bilayer, i.e. a homogeneous
solution, is denoted $\Omega_0(T,V,\zeta_1,\zeta_2,\zeta_s)$. The
difference between these two free energies, in the thermodynamic limit
of infinite volume, defines the excess free energy of the system with 
one, or more, membrane:
\begin{equation}
 \delta\Omega(T,{\cal A},\zeta_1,\zeta_2,\zeta_s)\equiv\lim_{V\rightarrow\infty}
[\Omega(T,V,{\cal A},\zeta_1,\zeta_2,\zeta_s)-\Omega_0(T,V,\zeta_1,\zeta_2,\zeta_s)].\end{equation}
With the excess free
energy known, the surface free energy per unit area, or equivalently,
the surface tension, $\gamma$, is obtained from the excess free energy of a single,
flat, bilayer $\delta\Omega_{bilayer}$
\begin{equation}
\gamma(T,\zeta_1,\zeta_2,\zeta_s)\equiv\lim_{{\cal A}\rightarrow\infty}
\frac{\delta\Omega_{bilayer}(T,{\cal A},\zeta_1,\zeta_2,\zeta_s)}{{\cal A}}.
\end{equation}

In order to calculate the free energy of stalk or hemifusion intermediates 
as a function of their radius, that radius must be fixed 
\cite{Katsov04,Matsen99} 
by a local Lagrange multiplier $\psi({\bf r}).$ Similarly, to constrain 
the membranes to be separated by a specified distance, $H$ at some point 
${\bf r}$, we must introduce an additional Lagrange multiplier, 
$\lambda({\bf r})$. The distance $H$ is chosen to be the distance between 
the hydrophilic/hydrophobic interfaces of the contacting, cis, leaflets as 
shown in Fig. 1.
With these additional constraints, the  free energy functional to be 
minimized now reads\begin{eqnarray}
\label{eq:systemenergy2}
\frac{Nv{\tilde\Omega}}{k_{B}T} &=& -\zeta_1Q_{1}-\zeta_{2}Q_{2}-\zeta_sQ_{s} 
+\int dV [\chi N \phi_{A}({\bf r})\phi_{B}({\bf r}) \nonumber \\
& & -w_{A}({\bf r})\phi_{A}({\bf r})-w_{B}({\bf r})\phi_{B}({\bf r}) 
-\xi({\bf r})(1-\phi_{A}({\bf r})-\phi_{B}({\bf r}))  \\
& & -\psi \delta(\rho-R)\delta(z)(\phi_{A}({\bf r})-\phi_{B}({\bf r}))  \\
& &
-\lambda [\delta(z-H/2)+\delta(z+H/2)](\phi_{A}({\bf r})-\phi_{B}({\bf r}))].\nonumber
\end{eqnarray} 
It is clear that one cannot constrain the bilayers to be a distance $H$ 
apart at a position at which the stalk or hemifusion diaphragm come in 
contact. Consequently in the last integral in Eq. (\ref{eq:systemenergy2}), 
the region of integration over $\rho$ is restricted to be greater than 
$R+R_c$, where $R$ is the radius of the fusion intermediate, and $R_c$ is 
positive and at least as large as the hydrophilic thickness of the 
bilayer. The condition that the free energy functional of Eq. 
(\ref{eq:systemenergy2}) be minimized yields a set of self-consistent 
equations that we solve in real space.  A detailed description on the 
derivation of Eq. (\ref{free1}) and the real space solution algorithm can 
be found elsewhere \cite{Katsov04,Fredrickson06,Mueller06}.

Finally we need to compare the energies in a biological system with those
in our homopolymer system. There are various choices for the energy of the biological
system. One could choose a property of a single bilayer, such as the energy per unit area of a hydrophobic, hydrophilic interface. Alternatively a property of two interacting bilayers coud be chosen, such as the attractive energy per unit area  between them. As the former is so well known, we shall employ it, but will show below that this gives essentially the same result had we chosen the latter.  We consider the dimensionless quantity 
$\gamma_{ow}D^2/k_BT$, where $\gamma_{ow}=40\times 10^{-3}$N/m is 
the oil, water interfacial
tension, and $D=4\times 10^{-9}$m is a typical bilayer thickness. With
$k_BT=4.3\times 10^{-21}$Nm, this ratio is about 150 for a biological system. 
The analogous quantity in the polymer system is $\gamma_0d^2/k_BT$,
where $\gamma_0$ is the surface tension between 
coexisting solutions of hydrophobic and 
hydrophilic homopolymers,
and $d$ is the thickness of our bilayers. We calculate
$\gamma_0d^2/k_BT=56.7$, so that energy scales in a biological system are
about a factor of 150/56.7=2.6 greater than in our polymer model.  

\section{Results and Discussion}
We first consider some properties of a single bilayer
composed of lamellar-forming amphiphiles, 
chosen to mimic DOPC, whose volume fraction is 0.6, and hexagonal-forming 
amphiphiles, chosen to mimic DOPE, whose volume fraction is 0.4. The 
leaves are of equal composition. As in our previous work, we have chosen the 
volume of amphiphile 1 to be $Nv=1.54 R_g^3$, where $R_g$ is the radius of 
gyration of the polymer. The bilayer thickness, measured between the 
planes at which the volume fractions of the hydrophilic part of the 
amphiphiles and that of the solvent are equal, is $4.3R_g$. The 
hydrophobic thickness, measured between the planes at which the volume 
fractions of hydrophilic and hydrophobic parts of the amphiphiles are 
equal, is $2.7 R_g.$

Two such bilayers have a weak attraction between them due to depletion 
forces induced by expulsion of some solvent when they are brought 
together. To see this, we calculate the excess free energy of a system 
of two flat bilayers a distance $H$ apart, $\delta\Omega_{2bilayers}(H)$ and define
the free energy per unit area  
\begin{equation}
\label{fofh}
F(H)\equiv \frac{\delta\Omega_{2bilayers}(H)}{{\cal A}}-2\gamma.
\end{equation}
By definition, this quantity asymptotes to zero for large $H$, and is negative 
when the bilayers attract one another. For the case of bilayers under
zero tension, the dimensionless quantity 
$F(H)R_g^2/k_BT$ is plotted in Fig. 2.

This energy of attraction per unit area can be compared with those measured between phospholipid bilayers provided we know the length scale given by $R_g$, the radius of gyration of the polymers in our system. To obtain this we note that the thickness of our bilayers is approximately 4.3$R_g$. If we take a typical bilayer thickness to be 4 nm, then $R_g\sim 0.93$ nm. With this and $k_BT=4.1\times 10^{-21}$ J, our calculated value of the free energy per unit area at the equilibrium distance between membranes corresponds to 0.07 mJ/m$^2$.  This should be increased by the factor of 2.6 if the energy scale we obtained by comparison with the hydrophilic, hydrophobic repulsion, is correct. Thus we expect that the energies of attraction per unit area between two phospholipid bilayers should be approximately 0.18 mJ/m$^2$.  This agrees extremely well with the results presented by Marra and Israelachvili \cite{Marra85} in their Fig. 2. It shows that we could have obtained  our energy scale equally well from the interaction energy of two bilayers.
 
 The excess free energy of an intermediate, such as a stalk, is calculated as follows.
 We compute the excess free energy, $\delta\Omega(H)$, of the system of two bilayers which are connected by the intermediate, and which, far from it, are separated by a distance $H$.
The excess free energy of the intermediate is, then
\begin{equation}
\delta\Omega_{int}(H)= \lim_{{\cal A}\rightarrow\infty}
\left\{\delta\Omega(H)-[F(H)+2\gamma]{\cal A}\right\}.
\end{equation}
In Figure 3, we show the excess surface energy of the stalk as a function 
of its radius, $R$, at different bilayer separations $H$.  Again, the
tension of the bilayer is zero. Each leaf of 
the bilayers shown here have compositions, $\phi_1=0.60$ and 
$\phi_2=0.40$, which are almost the same as the cis leaves of the 
asymmetric membranes we considered previously \cite{Lee07}.
We note that for stalk radii which
are quite small, less than about 0.5 $R_g$, we find no solution 
for a stalk-intermediate. This reflects the fact that the process by which the 
stalk initially forms cannot necessarily be thought of as one which produces a stalk of 
infinitesimal radius which then expands. At large radii, the stalk  expands into a hemifusion diaphragm. We find that as the membrane separation $H$ increases, this hemifusion diaphragm becomes 
indistinguishable from a single bilayer membrane. Hence for all large
$H$ the free energy increases linearly with $R$ with a slope directly
related to a line tension, one which arises from the junction of the
hemifusion diaphragm with the two bilayer membranes. 

The most important result in Fig. 3 is that the increase of separation
between fusing bilayers causes the energy of the
metastable stalk to increase significantly. It follows that the barrier
to the formation of this stalk also 
increases significantly with separation. As an estimate to this barrier, we 
take the energy of the stalk with the smallest radius for which we find
a solution of our equations. This should be considered an upper bound,
as there may be less expensive paths to the creation of the stalk.
A second result of note is that there is no metastable 
stalk if the bilayers are too close to one another. This is because the
energy associated with the rearrangement of amphiphiles needed to make
the stalk is simply too large at small membrane separations.
As the intermembrane distance increases, the stalk does become
metastable with a radius on the order of 1.3$R_g$. This is reasonable as
the diameter of this stalk is about the same as the hydrophobic
thickness of our bilayers, $2.7R_g$, so that 
amphiphiles that make 
up the stalk can take configurations somewhat similar to
those of amphiphiles in the 
unperturbed bilayers.

The importance of the stalk being metastable can be seen in Fig. 4, which 
summarizes the results of our calculation. We have plotted, as a function 
of separation, $H$, the energy of the stalk with the smallest radius
for which we find a solution, (squares), the 
energy of the metastable stalk (circles), and the barrier 
(triangles) which is associated with the expansion of the stalk into a 
hemifusion diaphragm before pore formation. For 
the smallest two interbilayer separations shown, $H=1.96 R_g$ and $H=2.20 
R_g$, there is no metastable stalk. Consequently one large activation 
energy of approximately $11 k_BT$ (corresponding to 29 $k_BT$ for a 
biological membrane) is required before a fusion pore can form.  However, 
for separations for which there is a metastable stalk, $H \geq 2.5 R_g$, fusion 
can occur in two steps: formation of the initial stalk which relaxes to 
the metastable stalk, and expansion into a hemifusion diaphragm with 
formation of a fusion pore. An additional activation energy is required 
for this second step, and is given by the difference between the energy of 
the second barrier and that of the metastable stalk. 
A third point of interest concerns the range from $H/R_g>2.49$, at
which the stalk first becomes metastable, to $H/R_g<3.05$ at which the
barrier to make the hemifusion diaphragm (triangles in Fig. 4) is no
longer larger than the barrier to make the initial stalk, (squares in
Fig. 4). Within this range, the
additional energy needed by the metastable stalk  
to surmount the second barrier and go forward to the
hemifusion diaphragm is larger than that required for the process to reverse
itself by means of the disappearance of the stalk. In other words, in
this range successful fusion is a less likely outcome of stalk formation 
than the simple disappearance of the stalk. The probabilities of these
outcomes are not reversed until $H/R_g$ exceeds 3.25. But at this larger
separation, the barrier to form the initial stalk is also larger. Thus
we expect that most of the time a metastable stalk actually forms, it
does not lead to successful fusion.   
 
A fourth point we wish to make is the following: 
once a metastable 
stalk becomes possible, the additional activation energy needed to pass
to the hemifusion diaphragm is 
always less than the barrier to create the initial stalk. Hence this 
barrier to create the initial stalk, whose magnitude is shown by the
squares in Fig. 4, becomes the 
largest barrier to fusion. Its magnitude is the smallest when the stalk 
first becomes metastable, which occurs when the bilayers are at a distance 
$H\sim 2.5R_g$ which exceeds by about 20\% a distance equal to half the
hydrophobic thickness of 
our bilayers. This small membrane separation, again defined between 
the hydrophilic/hydrophobic interfaces of the apposed cis leaflets can
be compared with 
the results of Weinreb and Lentz
\cite{Weinreb07} who found optimum fusion at a distance between
hydrophobic/hydrophilic interfaces that was comparable to half the
hydrophobic thickness of their bilayers. 
The value of this smallest barrier for stalk formation is, from Fig. 4, 
about 5.6$k_BT$ for the 
copolymer membranes, which corresponds to about 14.6 $k_BT$ for a biological 
membrane.

We note from Fig. 4 that the free energies of the metastable stalk and
of the barrier to its creation become linear functions of $H$ even for
values of $H$ which are not too large. The rate of increase of the free 
energy of the metastable stalk with 
intermembrane distance, (circles in Fig. 4), is 2.5 $k_BT/(H/R_g)$. 
We can convert this rate of change of free energy with distance to
practical units as follows. We increase the energy by a factor of 2.6 to
account for the difference between our amphiphilic bilayers and those
composed of lipids and utilize the length scale $R_g\sim 0.93$ nm obtained earlier. From these we find that the above rate of increase of the
metastable stalk free energy with thickness becomes
\begin{equation}
\frac{d\delta\Omega_{stalk}}{dH}\sim 7.1\ k_BT/{\rm nm},
\end{equation} 
As we have set the tension of the bilayers to zero, this increase in
stalk free energy 
does not arise simply from the additional surface area of a longer stalk.
We have repeated our calculations taking a surface tension equivalent to 
2.68 mN/m, a value in the range of tensions which can cause rupture \cite{Evans03},
and found that the rate of change of metastable stalk energy with membrane separation 
increased from the value of 7.1 k$_B$T/nm only to 9.4 k$_B$T/nm.  
Therefore we conclude that the increased area associated with a 
stalk of greater length is not the major  contribution to the stalk free energy. Rather	
it is plausible that the dominant contribution to the length dependence
of the metastable stalk free energy comes from the packing of the
hydrophobic tails. That is, although the stalk has a diameter comparable
to the hydrophobic thickness of the bilayer, the axially symmetric
configuration is very different from the planar bilayer. If the density
of headgroups in the stalk is comparable to that in the bilayer, then
the tails become  crowded near the center of the
stalk. Conversely, if the tail density at the center is comparable to
that of the interior of the bilayer, then the density of headgroups must
be considerably less than that of the bilayer causing a significant
energy penalty of contact between solvent and tails. 
This conjecture is strengthened by the observation,
from Fig. 4, that the rate of increase with distance, $H$, of the
barrier to stalk formation is greater than that for the metastable
stalk itself. This is reasonable as the intermediate that we consider,
and which corresponds to
the barrier, is a stalk of diameter smaller than that of the metastable
stalk, and also smaller than the thickness of an unperturbed bilayer. 
Hence the hydrophobic tails are packed quite densely.
From Fig. 4, (squares), this slope is  $d\delta\Omega_{barrier}=6.0\ d(H/R_g)$ which
for a biological membrane translates to    
\begin{equation}
\frac{d\delta\Omega_{barrier}}{d H}\sim 16.7\ k_BT/{\rm nm}.
\end{equation}

These results permit us to discuss the interesting case which arises    
when the apposing membranes cannot be
brought to the optimum, small, distance which the amphiphiles would
like simply because of the very size of the protein machinery which
brings the membranes together. This is the case with hemagglutinin whose
approximate 4nm width \cite{Wilson81} keeps the head groups of apposing membranes
this distance apart. If we assume a headgroup of 1 nm \cite{Marra85},
then the minimum
distance between hydrophilic/hydrophobic interfaces is on the order of 6nm. The
free energy of the metastable stalk and the barrier to its creation 
when the apposing bilayers are constrained to be at such a distance 
can be estimated from Fig. 4 and the linear behavior at large distances
given above.
We find the metastable stalk to have an excess free energy of 38
$k_BT$. The barrier to be overcome to create this metastable stalk is
about 73 $k_BT$. It is understandable that more than a single
hemagglutinin molecule is required to bring about the amphiphile
reorganization needed to produce a stalk linking membranes at such a
distance, one imposed by the very machinery of fusion itself.

\section{Acknowledgments} This work was supported by the National
Science Foundation under Grant No. DMR-0503752.   

\clearpage
\begin{figure}
   \begin{center}
   \includegraphics*[width=5.25in]{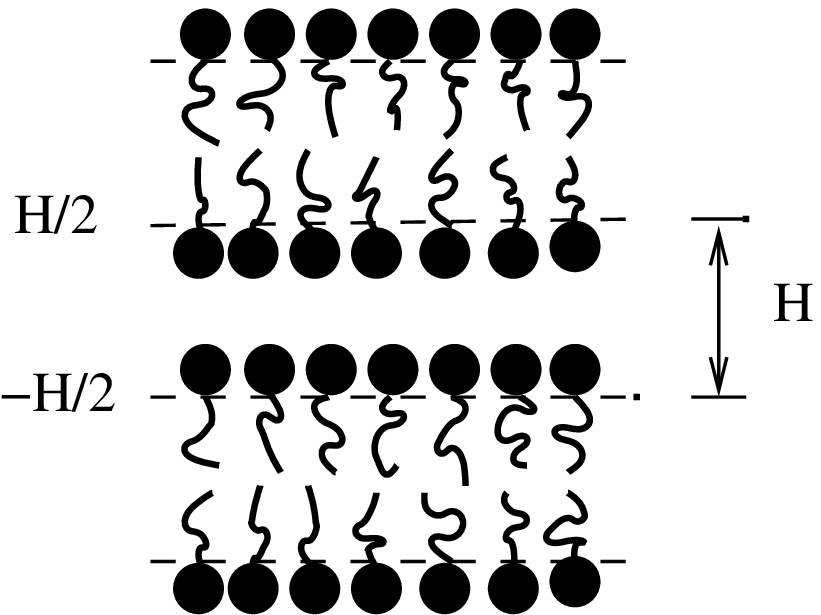}
      \caption{}
      \label{distance}
   \end{center}
\end{figure}
\clearpage
\begin{figure}
   \begin{center}
   \includegraphics*[width=5.25in]{fig2.eps}
      \caption{}
   \end{center}
\end{figure}
\clearpage
\begin{figure}
   \begin{center}
   \includegraphics*[width=5.25in]{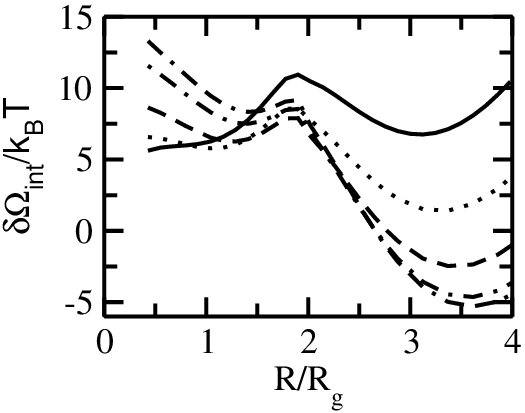}
      \caption{}
   \end{center}
\end{figure}
\clearpage
\begin{figure}
   \begin{center}
   \includegraphics*[width=5.25in]{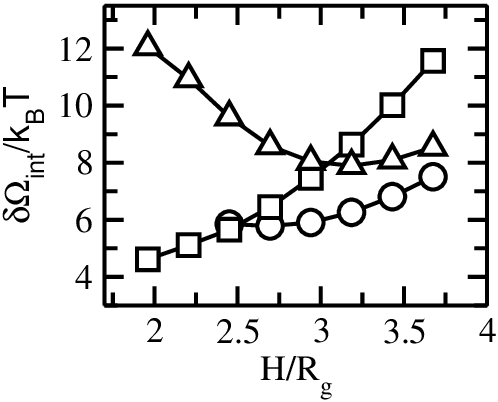}
      \caption{}
   \end{center}
\end{figure}
\clearpage
\section{Figure Captions}
\begin{itemize}
\item[]{Figure 1} Apposed bilayers separated by distance $H$. Circles 
represent hydrophobic head groups and curved lines hydrophobic tails.  
The separation $H$ is measured between the hydrophilic/hydrophobic 
interfaces of the contacting leaflets.
\item[]{Figure 2} Free energy per unit area 
of apposed bilayers, $F(H)$ of eq (\ref{fofh}), in units of $k_BT/R_g^2$ 
as a function of 
separation distance, $H/R_g$, between bilayers 
composed of 60\% lamellar formers 
and 40\% hexagonal formers and 
under zero tension.
\item[]{Figure 3} Excess surface energy of stalk-like fusion intermediates 
as a function of stalk radius, $R$ for $H=2.2 R_g$ (solid), $H=2.7 R_g$ 
(dotted), $H=3.2 R_g$ (dashed), $H=3.7 R_g$ (dot-dashed), and $H=4.0
R_g$ (dot double-dashed) for 
systems composed of 60\% DOPC-like and 40\% DOPE-like diblocks under 
zero tension. 
\item[]{Figure 4} Various energies related to fusion in the standard 
mechanism as a function of separation $H$ for bilayers shown in Fig. 3.  
Squares represent the initial barrier to create a stalk, circles the 
metastable stalk energy, and triangles the second barrier as the stalk expands 
to a hemifusion diaphragm.  For the lowest two values of separation ($H=1.96 
R_g$ and $2.20 R_g$), metastable stalks do not exist. 
\end{itemize}

\clearpage
\bibliography{014731JCP}
\end{document}